\documentclass[aps,pra,twocolumn,epsf,showpacs,superscriptaddress]{revtex4}
\usepackage{graphicx}
\usepackage{amsmath,amssymb,latexsym}
\usepackage{bm}

\begin{document}

\title{Linear dynamics of quantum-classical hybrids} 

\author{Hans-Thomas Elze}
\affiliation{Dipartimento di Fisica ``Enrico Fermi'',  
        Largo Pontecorvo 3, I-56127 Pisa, Italia } 

\email{elze@df.unipi.it}

\begin{abstract} 
A formulation of quantum-classical hybrid dynamics is presented, 
which concerns the direct  
coupling of classical and quantum mechanical degrees of freedom. It  
is of interest for applications in quantum mechanical approximation schemes 
and may be relevant for the foundations of quantum mechanics, in particular, when it 
comes to experiments exploring the quantum-classical border. The present linear theory 
differs from the nonlinear ensemble theory of Hall and Reginatto, but shares with it 
to fulfil all consistency requirements discussed in the literature, while earlier 
attempts failed in this respect. Our work is based on the representation of 
quantum mechanics in the framework of classical analytical mechanics by A.~Heslot, 
showing that notions of states in phase space, observables, Poisson brackets, 
and related canonical transformations  
can be naturally extended to quantum mechanics. This is suitably generalized for 
quantum-classical hybrids here. 
\end{abstract}
\maketitle 

\section{Introduction} 
The hypothetical direct coupling of quantum mechanical and classical degrees of 
freedom -- {\it ``hybrid dynamics''} -- presents a departure from quantum mechanics 
that has been researched through decades for practical as well as theoretical reasons. 
In particular, the standard Copenhagen interpretation has led to the unresolved 
measurement problem which, together with the fact that quantum mechanics needs  
such interpretation, in order to be operationally well defined, may indicate that 
it deserves amendments. In this context, it has been recognized early on that 
a theory which dynamically bridges the quantum-classical divide should have an 
impact on the measurement problem \cite{Sudarshan123}, besides being essential 
for attempts to describe consistently the interaction between quantum matter and classical 
spacetime \cite{BoucherTraschen}.  
   
Numerous works have appeared, in order to formulate hybrid dynamics in a satisfactory 
way. However, they were generally found to be deficient for one or another reason. 
This has led to various no-go theorems accompanying a list of desirable properties 
or consistency requirements, see, for example, 
Refs.\,\cite{CaroSalcedo99,DiosiGisinStrunz}:  
\begin{itemize} 
\item Conservation of energy. 
\item Conservation and positivity of probability. 
\item Separability of quantum and classical subsystems in the absence of their interaction, 
recovering the correct quantum and classical equations of motion. 
\item Consistent definitions of states and observables; existence of a Lie bracket structure 
on the algebra of observables that suitably generalizes  
Poisson and commutator brackets. 
\item Existence of canonical transformations generated by the observables; 
invariance of the classical sector under canonical transformations 
performed on the quantum sector only and {\it vice versa}. 
\item Existence of generalized Ehrenfest relations ({\it i.e.} the 
correspondence limit) which, 
for bilinearly coupled classical and quantum oscillators,  
are to assume the form of the classical equations of motion  
(``Peres-Terno benchmark'' test \cite{PeresTerno}). 
\end{itemize} 

These issues have been reviewed in recent works by Hall and Reginatto. Furthermore, 
there, they have introduced the first viable theory of hybrid dynamics that 
agrees with all points listed above \cite{HallReginatto05,Hall08,ReginattoHall08}. 
Their ensemble theory is based on configuration space, which entails a certain 
nonlinearity of the action functional from which it is derived. This nonlinearity 
leads to effects and a poposal to possibly falsify the theory experimentally 
\cite{HallEtAll11}. We will comment on this issue in due course (Subsection~5.3.).   
  
In fact, the aim of the present paper is to set up an alternative theory 
of hybrid dynamics, which is based on notions of phase space. 
This is partly motivated by work on related topics of general linear dynamics and 
classical path integrals \cite{EGV11,EGV10}.  
Presently, 
we will extend the work of Heslot, who has demonstrated that quantum 
mechanics can entirely be rephrased in the language and formalism of classical 
analytical mechanics \cite{Heslot85}. We thus introduce unified notions 
of states on phase space, observables, canonical transformations, and a generalized 
quantum-classical Poisson bracket in particular. This will lead to an intrinsically 
linear hybrid theory, which fulfils all consistency requirements as well.  

It may be worth while to comment on the 
relevance of hybrid dynamics, even if one is {\it not} inclined to modify     
certain ingredients of quantum theory. There is clearly practical 
interest in various forms of hybrid dynamics, in particular in nuclear, atomic, or 
molecular physics. The Born-Oppenheimer approximation, for example, is based on 
a separation of interacting slow and fast degrees 
of freedom of a compound object. The former are treated as approximately classical 
while the latter as of quantum mechanical nature. Furthermore, mean field theory, 
based on the expansion of quantum mechanical variables 
into a classical part plus quantum fluctuations, leads to another  
approximation scheme and another form of hybrid dynamics. This has been reviewed  
more generally for macroscopic quantum phenomena in Ref.\,\cite{Huetal}.  

In all these cases
hybrid dynamics is considered as an {\it approximate description} of  
an intrinsically quantum mechanical object. Which can lead to new insights, 
for example, to an alternative derivation of geometric forces and Berry's phase 
\cite{ZhangWu06}. 

Such considerations are and will become increasingly important for  
the precise manipulation of 
quantum mechanical objects by apparently and for all practical purposes classical means, 
especially in the mesoscopic regime. 
 
Furthermore, the   
backreaction effect of quantum fluctuations on classical degrees of freedom might be 
of considerable importance, 
in particular, if they originate in physically distinct ways. We recall continuing    
discussions of the ``semiclassical'' Einstein equation coupling the classical metric 
of spacetime to the expectation value of the energy-momentum tensor of quantized matter 
fields. Can this be made into a consistent hybrid theory leaving gravity unquantized? 
This has recently been re-examined, for example, in 
Refs.\,\cite{HallReginatto05,ReginattoHall08,Diosi10,HuetalGravity08}; 
various related aspects 
have been discussed, for example, in   
Refs.\,\cite{Diosi84,Penrose98,DiosiRev05,Adler03,Mavromatosetal92,Pullinetal08,Diosi09}. 
 
Finally, concerning speculative ideas about the emergence of quantum mechanics from a 
coarse-grained deterministic dynamics (see, for example, Refs.\,\cite{tHooft10,Elze09a,Adler} 
with numerous references to earlier work) the backreaction problem can be more 
provocatively stated as the problem of the interplay of fluctuations among underlying 
deterministic and emergent quantum mechanical degrees of freedom. Or, to put it differently:  
{\it ``Can quantum mechanics be seeded?''}

The remainder of the paper is organized as follows.
In Section~2., the results of Heslot's work are represented, in order 
to make the paper selfcontained, and we shall frequently refer to it in what follows. 
In Section~3., we introduce hybrid phase space ensembles and, in particular, the 
quantum-classical Poisson bracket which is central to our approach; the important 
issue of separability is resolved and time evolution discussed. In Section~4., 
hybrid dynamics is studied, incorporating quantum-classical interaction. Energy conservation, 
Ehrenfest relations, especially for bilinearly coupled oscillators, are derived there. 
In Section~5., we discuss various aspects of the present theory, in particular, the 
possibility to have classical-environment induced decoherence, the quantum-classical 
backreaction, a deviation from the Hall-Reginatto proposal predicted by 
the hybrid dynamics developed in this paper, and the closure 
of the algebra of hybrid observables. Section~6. presents  
concluding remarks.  

\section{Hamiltonian dynamics revisited}
In the following two subsections, we will briefly present some important  
results drawn from the remarkably clear exposition of classical Hamiltonian 
mechanics and its generalization incorporating quantum mechanics 
by Heslot \cite{Heslot85}. 
This will form the starting point of our discussion of the hypothetical 
direct coupling between quantum and classical degrees of freedom 
in Section~3. 

\subsection{Classical mechanics} 
The evolution of a {\it classical} object is described with respect to its $2n$-dimensional 
phase space, which is identified as its {\it state space}. A real-valued regular 
function on the state space defines an {\it observable}, {\it i.e.}, a differentiable function 
on this smooth manifold. 

Darboux's theorem shows that there always exist (local) systems of so-called 
{\it canonical coordinates}, commonly denoted by $(x_k,p_k),\; k=1,\dots ,n$, 
such that the {\it Poisson bracket} of any pair of observables $f,g$ assumes 
the standard form \cite{Arnold}: 
\begin{equation}\label{PoissonBracket} 
\{ f,g \}\; =\; 
\sum_k\Big (\frac{\partial f}{\partial_{x_k}}\frac{\partial g}{\partial_{p_k}}
-\frac{\partial f}{\partial_{p_k}}\frac{\partial g}{\partial_{x_k}}\Big ) 
\;\;. \end{equation} 
This is consistent with $\{ x_k,p_l\}=\delta_{kl}$, $\{ x_k,x_l\} =\{ p_k,p_l\}=0,\; 
k,l=1,\dots ,n$, and reflects the bilinearity, antisymmetry, derivation-like 
product formula, and Jacobi identity which define a Lie bracket operation, 
$f,g\rightarrow\{ f,g\}$, mapping a pair of observables to an observable.  

Compatibility with the Poisson bracket structure restricts general transformations 
${\cal G}$ of the state space to so-called {\it canonical transformations} 
which do not change the physical properties of the object under study; e.g., a translation, 
a rotation, a change of inertial frame, or evolution in time. Such ${\cal G}$ 
induces a change of an observable, $f\rightarrow {\cal G}(f)$, and is an automorphism 
of the state space compatible with its Poisson bracket structure, if and only if, 
for any pair of observables $f,g$: 
\begin{equation}\label{canTrans} 
{\cal G}(\{ f,g\} )=\{ {\cal G}(f),{\cal G}(g)\} 
\;\;. \end{equation} 

Due to the Lie group structure of the set of canonical transformations, it is sufficient 
to consider infinitesimal transformations generated by the elements of the 
corresponding Lie algebra. Then, an {\it infinitesimal transformation} ${\cal G}$ is 
{\it canonical}, if and only if for any observable $f$ the map $f\rightarrow {\cal G}(f)$ 
is given by $f\rightarrow f'=f+\{ f,g\}\delta\alpha$, with some observable $g$, 
the so-called {\it generator} of ${\cal G}$, and $\delta\alpha$ an infinitesimal real 
parameter. 

Thus, for the canonical coordinates, in particular, an infinitesimal canonical transformation 
amounts to: 
\begin{eqnarray}\label{xcan} 
x_k&\rightarrow &x_k'=x_k+\frac{\partial g}{\partial p_k}\delta\alpha 
\;\;, \\ [1ex] \label{pcan}  
p_k&\rightarrow &p_k'=p_k-\frac{\partial g}{\partial x_k}\delta\alpha 
\;\;, \end{eqnarray} 
employing the Poisson bracket given in Eq.\,(\ref{PoissonBracket}). 

This analysis shows the fundamental relation between observables and generators 
of infinitesimal canonical transformations in classical Hamiltonian mechanics. 

For example, the energy ${\cal H}_{\mbox{\scriptsize CL}}$ given by the classical 
Hamiltonian function is the generator of time evolution: for 
$g={\cal H}_{\mbox{\scriptsize CL}}$ and $\delta\alpha =\delta t$, 
the Eqs.\,(\ref{xcan}) and (\ref{pcan}) are equivalent to Hamilton's equations, 
considering an infinitesimal time step $\delta t$. 
 
\subsection{Quantum mechanics} 
An important achievement of Heslot's work is the realization that the analysis summarized 
in the previous subsection can be generalized and applied to quantum mechanics; 
in particular, the dynamical aspects of quantum mechanics thus find a description 
in classical terms borrowed from Hamiltonian mechanics. Some related ideas have been 
presented earlier in Ref.\,\cite{Strocchi}.   

\subsubsection{Preliminaries}
To begin with, we recall that the Schr\"odinger equation and its adjoint can be 
obtained by requiring the variation with respect to state vector $|\Psi\rangle$ and 
adjoint state vector  $\langle\Psi |$, respectively, of the following action $S$ to vanish: 
\begin{equation}\label{action} 
S:=\int\mbox{d}t\;\langle\Psi (t)|(i\partial_t-\hat H)|\Psi (t)\rangle \equiv\int\mbox{d}t\; L 
\;\;, \end{equation} 
which involves the self-adjoint Hamilton operator $\hat H$ pertaining to the physical 
object under study. -- The adjoint equation follows after a partial integration, 
provided the surface terms do not contribute. This is guaranteed by the {\it normalization} 
condition: 
\begin{equation}\label{normalization} 
\langle\Psi (t)|\Psi (t)\rangle\stackrel{!}{=}\mbox{constant}\equiv 1 
\;\;, \end{equation} 
which is an essential ingredient of the probability interpretation associated with 
state vectors. -- Adding here that state vectors that differ by an {\it unphysical constant 
phase} are to be identified, we recover that the {\it quantum mechanical state space} 
is formed by the rays of the underlying Hilbert space, {\it i.e.}, forming a complex 
projective space. 

Making use of the Lagrangian $L$, defined as the integrand of the above action $S$, 
we define a momentum conjugate to the state vector: 
\begin{equation}\label{Pi} 
\langle\Pi |:=\frac{\partial L}{\partial |\dot\Psi\rangle} =i\langle\Psi | 
\;\;, \end{equation} 
with $|\dot\Psi\rangle :=\partial_t|\Psi\rangle$, and obtain the   
corresponding Hamiltonian function: 
\begin{equation}\label{HamiltonianQM}
\langle\Pi |\dot\Psi\rangle -L=-i\langle\Pi |\hat H|\Psi\rangle =:{\cal H}(\Pi ,\Psi ) 
\;\;. \end{equation} 
Finally, considering Hamilton's equations, deriving from ${\cal H}$: 
\begin{eqnarray}\label{Hamilton1} 
\partial_t|\Psi\rangle &=&\frac{\partial {\cal H}}{\partial\langle\Pi |}=-i\hat H|\Psi\rangle 
\;\;, \\ [1ex] \label{Hamilton2} 
\partial_t\langle\Pi |&=&-\frac{\partial {\cal H}}{\partial |\Psi\rangle}=
i\langle\Pi |\hat H 
\;\;, \end{eqnarray} 
we see indeed that they represent Schr\"odinger's equation and its adjoint, using  
$\langle\Pi |=i\langle\Psi |$, keeping the essential  
normalization condition (\ref{normalization}) in mind.  

\subsubsection{The oscillator representation}
Quantum mechanical evolution can be described by a unitary transformation, 
$|\Psi (t)\rangle =\hat U(t-t_0)|\Psi (t_0)\rangle$, with $U(t-t_0)=\exp [-i\hat H(t-t_0)]$, 
which formally solves the Schr\"odinger equation. 
It follows immediately that a stationary state, characterized by 
$\hat H|\phi_i\rangle =E_i|\phi_i\rangle$, with a real energy eigenvalue $E_i$,  
performs a simple harmonic motion, {\it i.e.}, 
$|\psi_i(t)\rangle =\exp [-iE_i(t-t_0)]|\psi_i(t_0)\rangle
\equiv\exp [-iE_i(t-t_0)]|\phi_i\rangle$. Henceforth, we assume a denumerable set of 
such eigenstates of the Hamilton operator.

Having recognized already the Hamiltonian character of the underlying 
equation(s) of motion, the harmonic motion suggests 
to introduce what may be called {\it oscillator representation} for such states. 
More generally, we consider 
the expansion of any state vector with respect to a complete 
orthonormal basis, $\{ |\Phi_i\rangle\}$:  
\begin{equation}\label{oscillexp} 
|\Psi\rangle =\sum_i|\Phi_i\rangle (X_i+iP_i)/\sqrt 2 
\;\;, \end{equation} 
where the generally time dependent expansion coefficients are explicitly written 
in terms of real and imaginary parts, $X_i,P_i$. Employing this expansion, allows to   
evaluate more explicitly the Hamiltonian function 
introduced in Eq.\,(\ref{HamiltonianQM}), {\it i.e.}, ${\cal H}=\langle\Psi |\hat H|\Psi\rangle$: 
\begin{eqnarray}
{\cal H}&=&\frac{1}{2}\sum_{i,j}\langle\Phi_i|\hat H|\Phi_j\rangle (X_i-iP_i)(X_j+iP_j) 
\nonumber \\ [1ex] \label{HamiltonianQM1}  
&=:&{\cal H}(X_i,P_i) 
\;\;. \end{eqnarray} 
Choosing especially the set of energy eigenstates, $\{ |\phi_i\rangle\}$, 
as basis for the expansion, we obtain:     
\begin{equation}\label{HamiltonianQM2} 
{\cal H}(X_i,P_i)=\sum_i\frac{E_i}{2}(P_i^{\;2}+X_i^{\;2}) 
\;\;, \end{equation} 
hence the name {\it oscillator representation}.  
The simple reasoning leading to this result clearly indicates that $(X_i,P_i)$ 
may play the role of {\it canonical coordinates} in the description of a quantum mechanical 
object and its evolution with respect to the state space. -- 
However, several points need to be clarified, in order to validate this interpretation. 

First of all, with $(X_i,P_i)$ as canonical coordinates and ${\cal H}$ as 
Hamiltonian function, we verify that the Schr\"odinger equation is recovered by 
evaluating $|\dot\Psi\rangle =\sum_i|\Phi_i\rangle (\dot X_i+i\dot P_i)/\sqrt 2$ 
according to Hamilton's equations of motion: 
\begin{eqnarray}
\dot X_i&=&\frac{\partial {\cal H}(X_j,P_j)}{\partial P_i}  
\nonumber \\ [1ex] \label{QHamiltonianEq1}
&=& -\frac{i}{2}\sum_j\Big (H_{ij}(X_j+iP_j)-(X_j-iP_j)H_{ji}\Big )
,\;\;\;  \\ [1ex]
\dot P_i&=&-\frac{\partial {\cal H}(X_j,P_j)}{\partial X_i}
\nonumber \\ [1ex]  \label{QHamiltonianEq2}  
 &=& -\frac{1}{2}\sum_j\Big (H_{ij}(X_j+iP_j)+(X_j-iP_j)H_{ji}\Big )
,\;\;\;  \end{eqnarray} 
where $H_{ij}:=\langle\Phi_i|\hat H|\Phi_j\rangle =H_{ji}^{\;\ast }$. Inserting these terms 
and using Eq.\,(\ref{oscillexp}) leads to $|\dot\Psi\rangle =-i\hat H|\Psi\rangle$, as expected. 
Using ${\cal H}$ in the special form given by Eq.\,(\ref{HamiltonianQM2}), we see that a zero 
mode with $E_{i'}=0$ automatically leads to $(X_{i'},P_{i'})=\mbox{constant}$.  

Secondly, the {\it constraint} ${\cal C}:=\langle\Psi |\Psi\rangle\stackrel{!}{=}1$, cf. 
Eq.\,(\ref{normalization}), becomes: 
\begin{equation}\label{oscillnormalization} 
{\cal C}(X_i,P_i)=\frac{1}{2}\sum_i(X_i^{\;2}+P_i^{\;2})\stackrel{!}{=}1 
\;\;. \end{equation} 
Thus, the vector with components given by the canonical coordinates 
$(X_i,P_i),\; i=1,\dots ,N$, is 
constrained to the surface of a $2N$-dimensional sphere with radius $\sqrt 2$. 
This constraint obviously presents a major difference to classical Hamiltonian mechanics. 

Following the previous discussion in Subsection~2.1., it is natural to introduce 
also here a {\it Poisson bracket} for any two {\it observables} on the 
{\it spherically compactified state space}, {\it i.e.} real-valued regular functions $F,G$ of 
the coordinates $(X_i,P_i)$:     
\begin{equation}\label{QMPoissonBracket} 
\{ F,G \}\; =\; 
\sum_i\Big (\frac{\partial F}{\partial_{X_i}}\frac{\partial G}{\partial_{P_i}}
-\frac{\partial F}{\partial_{P_i}}\frac{\partial G}{\partial_{X_i}}\Big ) 
\;\;, \end{equation} 
cf. Eq.\,(\ref{PoissonBracket}). -- Then, as before, the Hamiltonian acts as the generator 
of time evolution of any observable $O$, {\it i.e.}: 
\begin{equation}\label{evolution}  
\frac{\mbox{d}O}{\mbox{d}t}=\partial_tO+\{ O,{\cal H} \} 
\;\;. \end{equation} 
In particular, it is straightforward to verify with the help of 
Eqs.\,(\ref{QHamiltonianEq1})--(\ref{QHamiltonianEq2}) that the constraint, 
Eq.\,(\ref{oscillnormalization}), is conserved under the Hamiltonian flow:  
\begin{equation}\label{constraint} 
\frac{\mbox{d}{\cal C}}{\mbox{d}t}=\{ {\cal C},{\cal H} \}=0 
\;\;. \end{equation} 
Therefore, it is sufficient to impose this constraint, which implements the normalization 
of the quantum mechanical state, on the initial condition of time evolution.  

It remains to demonstrate the {\it compatibility} of the notion of 
{\it observable} introduced here -- as in classical mechanics, cf. the 
discussion leading to Eq.\,(\ref{canTrans}) and thereafter -- with the one adopted in 
quantum mechanics. This concerns, in particular, the implementation of {\it canonical 
transformations} and the role of observables as their generators.    

\subsubsection{Canonical transformations and quantum observables} 
The Hamiltonian function has been introduced as observable in the 
Eq.\,(\ref{HamiltonianQM1}) which provides a direct relation to the corresponding quantum  
observable, namely the expectation value of the self-adjoint Hamilton operator. 
This is an indication of the general structure to be discussed now.    

Refering to Section~III. of Heslot's work \cite{Heslot85} for details of the 
derivations, we summarize here the main points, which will be useful in the following:  
\\ \noindent $\bullet$ 
A) {\it Compatibility of unitary transformations and Poisson structure.} -- 
The canonical transformations discussed in Section~2.1. represent 
automorphisms of the classical state space which are compatible with the Poisson brackets. 
In quantum mechanics automorphisms of the Hilbert space are implemented by 
unitary transformations, $|\Psi '\rangle =\hat U|\Psi\rangle$, with $\hat U\hat U^\dagger =
\hat U^\dagger \hat U=1$. This implies a transformation of the canonical 
coordinates here, {\it i.e.}, of the expansion coefficients $(X_i,P_i)$ introduced in 
Eq.\,(\ref{oscillexp}): 
\begin{eqnarray}
|\Psi '\rangle &=&\sum_{i,j}|\Phi_i\rangle\langle\Phi_i|\hat U|\Phi_j\rangle 
\frac{X_j+iP_j}{\sqrt 2} 
\nonumber \\ [1ex] \label{oscillexptransf}   
&=&\sum_i|\Phi_i\rangle\frac{X_i'+iP_i'}{\sqrt 2}
\;\;. \end{eqnarray}  
Splitting the matrix elements $U_{ij}:=\langle\Phi_i|\hat U|\Phi_j\rangle$ into real and 
imaginary parts and separating Eq.\,(\ref{oscillexptransf}) accordingly, using  
orthonormality of the basis, yields the transformed coordinates in terms of the 
original ones. Then, a simple calculation, employing the 
Poisson bracket defined in Eq.\,(\ref{QMPoissonBracket}), shows that 
$\{ X_i',P_j'\} =\delta_{ij}$ and $\{ X_i',X_j'\} =\{ P_i',P_j'\} =0$, as before. The 
fundamental Poisson brackets remain invariant under unitary transformations. More generally, 
this implies \cite{Arnold} that $U(\{ F,G\}) =\{ U(F),U(G)\}$, cf. Eq.\,(\ref{canTrans}). 
Thus, {\it unitary transformations on Hilbert space are canonical transformations 
on the $(X,P)$ state space}.   
\\ \noindent $\bullet$ 
B) {\it Self-adjoint operators as observables.} -- 
Any infinitesimal unitary transformation $\hat U$ can be generated by a self-adjoint operator 
$\hat G$, such that: 
\begin{equation}\label{Uinfini} 
\hat U=1-i\hat G\delta\alpha 
\;\;, \end{equation} 
which will lead to the quantum mechanical relation between an observable and 
a self-adjoint operator, replacing the classical construction in Section~2.1. In fact, 
straightforward calculation along the lines of A), splitting matrix elements of $\hat G$, 
with $G_{ji}^\ast =G_{ij}$, into real and imaginary parts, shows that in the present case 
we have: 
\begin{eqnarray}\label{Xcan} 
X_i&\rightarrow &X_i'=X_i+\frac{\partial \langle\Psi |\hat G|\Psi\rangle}
{\partial P_i}\delta\alpha 
\;\;, \\ [1ex] \label{Pcan}  
P_i&\rightarrow &P_i'=P_i-\frac{\partial \langle\Psi |\hat G|\Psi\rangle}
{\partial X_i}\delta\alpha 
\;\;. \end{eqnarray} 
Due to the phase arbitrariness -- $\hat U$ and $\hat U\cdot\exp (i\theta )$, with constant 
phase $\theta$, are physically equivalent -- the operator $\hat G$ is defined up 
to an additive constant. This constant is naturally chosen such that the  
relation between an observable $G$, defined in analogy to Section~2.1., and a self-adjoint 
operator $\hat G$ can be inferred from Eqs.\,(\ref{Xcan})--(\ref{Pcan}): 
\begin{equation}\label{goperator} 
G(X_i,P_i)=\langle\Psi |\hat G|\Psi\rangle 
\;\;, \end{equation} 
by comparison with the classical result, Eqs.\,(\ref{xcan})--(\ref{pcan}).  
In conclusion, a {\it real-valued regular function $G$ of the state is an observable, if 
and only if there exists a self-adjoint operator $\hat G$ such that Eq.\,(\ref{goperator}) 
holds}. -- 
Note that {\it all quantum observables are quadratic forms} 
in the $X_i$'s and $P_i$'s. This explains that there are much fewer of them  
than in the corresponding classical case. 
\\ \noindent $\bullet$ 
C) {\it Commutators as Poisson brackets.} -- 
The relation (\ref{goperator}) between observables and self-adjoint operators is linear 
and admits $\hat{\bm 1}$ as unit operator, since 
$\langle\Psi |\Psi\rangle\stackrel{!}{=}1$. Therefore, addition of observables and 
multiplication by a scalar of observables are well-defined and translate into the 
corresponding expressions for the operators. One may then consider the 
Poisson bracket (\ref{QMPoissonBracket}) of two observables and 
demonstrate the important result \cite{Heslot85}: 
\begin{equation}\label{QMPBComm} 
\{ F,G\}=\langle\Psi |\frac{1}{i}[\hat F,\hat G]|\Psi\rangle 
\;\;, \end{equation}  
with both sides of the equality considered as functions of the variables $X_i,P_i$, 
of course, and with the commutator defined as usual, 
$[\hat F,\hat G]:=\hat F\hat G-\hat G\hat F$. This shows that 
the {\it commutator is a Poisson bracket with respect to the $(X,P)$ state space} and 
relates the algebra of observables, in the sense of the classical construction of Section~2.1.,  
to the algebra of self-adjoint operators in quantum mechanics.   
\\ \noindent $\bullet$ 
D) {\it Normalization, phase arbitrariness, and admissible observables.} -- 
Coming back to the normalization condition $\langle\Psi |\Psi\rangle\stackrel{!}{=}1$, 
which compactifies the state space, cf. the constraint Eq.\,(\ref{oscillnormalization}), 
it must be preserved under infinitesimal canonical transformations, since it belongs 
to the structural characteristics of the state space. By Eqs.\,(\ref{Xcan})--(\ref{goperator}), 
an infinitesimal canonical transformation generated by an observable $G$ leads to 
${\cal C}(X_i,P_i)\;\rightarrow\; {\cal C}(X_i',P_i')$, with: 
\begin{eqnarray}
{\cal C}(X_i',P_i')&=&{\cal C}(X_i,P_i) 
+\sum_j\Big (
\frac{\partial G}{\partial P_j}X_j-\frac{\partial G}{\partial X_j}P_j\Big )\delta\alpha 
\nonumber \\ [1ex] \label{constraintinvar} 
&\;&
+\mbox{O}(\delta\alpha^2)
\;\;. \end{eqnarray} 
Therefore, a {\it necessary} condition which observables must fulfil is the vanishing of 
the term $\propto\delta\alpha$ here, {\it i.e.} the {\it invariance of the constraint} under 
such transformations, ${\cal C}(X_i',P_i')\stackrel{!}{=}{\cal C}(X_i,P_i)$. 
It is {\it not sufficient}, since the product $G_1G_2$ of two observables -- which fulfil 
this condition individually and, therefore, their product as well -- does not necessarily represent an observable: the corresponding  
self-adjoint operators do not necessarily commute, 
{\it i.e.}, generally we have $(\hat G_1\hat G_2)^\dagger =G_2G_1\neq G_1G_2$. 
-- Incidentally, the condition of the vanishing second term on the right-hand side 
of Eq.\,(\ref{constraintinvar}) follows also 
more generally, via Eq.\,(\ref{goperator}), from the requirement that 
any observable $G$ is invariant under an infinitesimal 
phase transformation $|\Psi\rangle\;\rightarrow\; |\Psi\rangle\cdot\exp (i\delta\theta )$, 
with constant $\delta\theta$, 
$G(X_i',P_i')\stackrel{!}{=}G(X_i,P_i)$. Conversely, assuming this {\it phase invariance} 
of the observables, we 
recover that Hilbert space vectors differing by an arbitrary constant phase are 
indistinguishable and represent the same physical state. 

We note that any observable $G$ with an expansion as in Eq.\,(\ref{HamiltonianQM1})  
automatically satisfies the invariance requirements of item D) above, 
the vanishing of the second term on the right-hand side of Eq.\,(\ref{constraintinvar}), 
in particular. Explicit calculation shows:  
\begin{equation}\label{constraintinvar1}
\{ {\cal C},G\}=\sum_j\Big (
\frac{\partial G}{\partial P_j}X_j-\frac{\partial G}{\partial X_j}P_j\Big )=0
\;\;, \end{equation} 
assuming that: 
\begin{equation}\label{Gexp} 
G(P_i,X_i):=\langle\Psi |\hat G|\Psi\rangle 
=\frac{1}{2}\sum_{i,j}G_{ij}(X_i-iP_i)(X_j+iP_j)
, \end{equation} 
and where $G_{ij}:=\langle\Phi_i|\hat G|\Phi_j\rangle =G_{ji}^\ast$, for a 
self-adjoint operator $\hat G$. 

In conclusion, quantum mechanics shares with classical mechanics an even dimensional state 
space, a Poisson structure, and a related algebra of observables. Yet it  
differs essentially by a restricted set of observables and the requirements 
of phase invariance and normalization, which compactify the underlying Hilbert space 
to the complex projective space formed by its rays.  

\section{Hybrid phase space ensembles} 
So far, we have described the Hamiltonian formalism of classical mechanics 
and its generalization which covers  
quantum mechanics, by adding more structure to the relevant state space. 
With the Hamiltonian equations of motion at hand, we could 
proceed to study the evolution and direct coupling of classical and 
quantum objects. However, it is convenient to study the evolution of ensembles over the 
state (or phase) space instead \cite{EGV11,EGV10}. Last not least, this will allow to include  
quantum mechanical mixed states, thus generalizing beyond the 
above tacitly assumed pure states.  

In this section, we still {\it neglect interactions between classical and 
quantum sectors} of a combined system, the study of which will lead us to truly 
quantum-classical hybrids only in the next Section~4. -- From now on, we will refer to 
the classical and quantum sectors as ``CL'' and ``QM'', respectively.    

We describe a {\it quantum-classical hybrid ensemble} by a real-valued,  
positive semi-definite, normalized, and possibly time dependent regular function, the 
{\it probability distribution} $\rho$, on the {\it Cartesian 
product state space} canonically coordinated by $2(n+N)$-tuples $(x_k,p_k;X_i,P_i)$; 
we reserve variables 
$x_k,p_k,\; k=1,\dots ,n$ for the CL sector (cf. Section~2.1.) and  
variables $X_i,P_i,\; i=1,\dots ,N$ for the QM sector (cf. Section~2.2.). 
One or the other sector of the state space may eventually be infinite dimensional. 
A physical realization of such an ensemble can be imagined as a collection of 
representatives of the combined system with different initial conditions.   

In order to qualify as {\it observable}, the distribution $\rho$ additionally has to 
obey the constraint induced by the extra structure of the QM sector of 
state space, see Subsection~2.2.3.~D). 
Evaluating the expectation of the corresponding self-adjoint, positive semi-definite, 
trace normalized {\it density operator} 
$\hat\rho$ in a generic state $|\Psi\rangle$, Eq.\,(\ref{oscillexp}), we have 
$\rho (x_k,p_k;X_i,P_i):=\langle\Psi |\hat\rho (x_k,p_k)|\Psi\rangle$, and: 
\begin{equation}\label{rhoexp} 
\rho (x_k,p_k;X_i,P_i) 
=\frac{1}{2}\sum_{i,j}\rho_{ij}(x_k,p_k)(X_i-iP_i)(X_j+iP_j)
, \end{equation} 
with $\rho_{ij}(x_k,p_k):=\langle\Phi_i|\hat\rho (x_k,p_k)|\Phi_j\rangle 
=\rho_{ji}^\ast (x_k,p_k)$.  
This assures that $\rho$ (or the  
marginal QM distribution obtained by integrating over the CL variables), as generator 
of a canonical transformation, 
does not violate the normalization constraint and phase invariance; in particular, 
it follows that $\{ {\cal C},\rho\} =0$, cf. Eq.\,(\ref{constraintinvar1}).  

Furthermore, {\it positive semi-definiteness} of $\rho$ imposes constraints   
on any other observable $G$ ($g$) of the QM (CL) sector, which can generate a 
canonical transformation. Considering infinitesimal transformations in 
both sectors, cf. Eqs.\,(\ref{xcan})--(\ref{pcan}) and Eqs.\,(\ref{Xcan})--(\ref{goperator}), 
we obtain $\rho (x_k,p_k;X_i,P_i)\;\rightarrow\;\rho (x_k',p_k';P_i',X_i')$, with: 
\begin{eqnarray}
\rho (x_k,p_k;X_i,P_i)&=&\rho (x_k,p_k;P_i,X_i)
\nonumber \\ [1ex] 
&\;&+(\partial_{x_k}\rho\;\partial_{p_k}g-\partial_{p_k}\rho\;\partial_{x_k}g)
\delta\alpha_{\mbox{\scriptsize CL}} 
\nonumber \\ [1ex] 
&\;&+(\partial_{X_i}\rho\;\partial_{P_i}G-\partial_{P_i}\rho\;\partial_{X_i}G)
\delta\alpha_{\mbox{\scriptsize QM}} 
\nonumber \\ [1ex] \label{rhotransf}
&\;&+\mbox{O}(\delta\alpha^{\;2}_{\mbox{\scriptsize CL}},
\delta\alpha^{\;2}_{\mbox{\scriptsize QM}}, 
\delta\alpha_{\mbox{\scriptsize CL}}\delta\alpha_{\mbox{\scriptsize QM}}) 
\;.\;\; \end{eqnarray} 
Now, if and where the distribution $\rho$ vanishes, also the first order terms 
$\propto\delta\alpha_{\mbox{\scriptsize CL}}$ and 
$\propto\delta\alpha_{\mbox{\scriptsize QM}}$ have to vanish, since otherwise  
$\rho$ can be made to decrease below zero by suitably choosing the signs of 
these independent infinitesimal parameters. 

This will be particularly relevant for the time evolution generated by 
a quantum-classical hybrid Hamiltonian, to be discussed in Section~4. 

\subsection{The probability density and marginal distributions}
Finally, we remark that the relation between an observable in $(X,P)$-space 
and a self-adjoint operator, Eq.\,(\ref{goperator}), can be written as: 
$G(X_i,P_i)=\mbox{Tr}(|\Psi\rangle\langle\Psi |\hat G)$,  
which shows explicitly the role of a QM pure state as one-dimensional projector, in this 
context. In order to illuminate the meaning of the probability density $\rho$, 
we may then use the representation of $\hat\rho$ in terms of its eigenstates,   
$\hat\rho =\sum_jw_j|j\rangle\langle j|$, and obtain: 
\begin{eqnarray}
\rho (x_k,p_k;X_i,P_i)&=&\sum_jw_j(x_k,p_k)\mbox{Tr}(|\Psi\rangle\langle\Psi |j\rangle\langle j|) 
\nonumber \\ [1ex] \label{rhointerpr}
&=&\sum_jw_j(x_k,p_k)|\langle j|\Psi\rangle |^2  
\;\;,  \end{eqnarray} 
with $0\leq w_j\leq 1$ and $\sum_j\int\Pi_l(\mbox{d}x_l\mbox{d}p_l)w_j(x_k,p_k)=1$. 
    
We see that $\rho (x_k,p_k;X_i,P_i)$, when properly  
normalized, is the probability density to find in the hybrid ensemble the QM state 
$|\Psi\rangle$, parametrized by $X_i,P_i$ through Eq.\,(\ref{oscillexp}), {\it together with} 
the CL state described by the coordinates $(x_k,p_k)$ of a point in CL phase space. 

The probability density allows to evaluate expectations of QM, CL, or hybrid 
observables in the usual way. Particularly useful are also the 
{\it marginal} (or {\it reduced}) distributions defined by: 
\begin{eqnarray}
&\;&\rho_{\mbox{\scriptsize CL}}(x_k,p_k)\;:=\;
\nonumber \\ [1ex] \label{CLreduced} 
&\;&\;\;\Gamma_N^{-1}\int_{\delta S_{2N}(\sqrt 2)}
\Pi_j(\mbox{d}X_j\mbox{d}P_j)\;\rho (x_k,p_k;X_i,P_i) 
\;, \\ [1ex] \label{QMreduced}  
&\;&\rho_{\mbox{\scriptsize QM}}(X_i,P_i)\;:=\;\int\Pi_l(\mbox{d}x_l\mbox{d}p_l)
\rho (x_k,p_k;X_i,P_i) 
,\;\; \;\end{eqnarray} 
where $\Gamma_N$ denotes a normalization factor, to be determined shortly, 
and the integration in Eq.\,(\ref{CLreduced}) extends over the surface of 
a $2N$-dimensional sphere of radius $\sqrt2$, in 
accordance with Eq.\,(\ref{oscillnormalization}); the integration in Eq.\,(\ref{QMreduced}) 
extends over all the state space of the CL subsystem. 
The convergence of the integrals is assured by the positive semi-definiteness and 
normalizability  of $\rho$; the underlying assumption is that the CL subsystem 
occupies essentially only a finite region of its phase space, while the QM subsystem 
is constrained by the normalization of its state.   

More explicitly, using the representation given in Eq.\,(\ref{rhointerpr}), 
we calculate for a state vector $|\Psi\rangle$, expanded according to Eq.\,(\ref{oscillexp}): 
\begin{equation}\label{CLreduced1}
\rho_{\mbox{\scriptsize CL}}(x_k,p_k)
=\sum_jw_j(x_k,p_k)\sum_{i_1,i_2}
\langle j|\Phi_{i_1}\rangle\langle\Phi_{i_2}|j\rangle\cdot I_{i_1i_2}
\;, \end{equation}
with a remaining surface integral defined by:   
\begin{eqnarray}
I_{ab}&:=&
\Gamma_N^{-1}\int_{\delta S_{2N}(\sqrt 2)}\Pi_c(\mbox{d}X_c\mbox{d}P_c)\; 
\nonumber \\ [1ex] \label{surfaceint}
&\;&\;\;\;\;\;\;\;\;\;\;\;\;\;\;\;\;\;  \times (X_a+iP_a)(X_b-iP_b) 
\;\;, \end{eqnarray} 
and evaluated as follows: 
\begin{eqnarray} 
I_{ab}&=&
\delta_{ab}\Gamma_N^{-1}\int\Pi_c(\mbox{d}X_c\mbox{d}P_c)\; 
\delta\Big( 2-\sum_i(X_i^{\;2}+P_i^{\;2})\Big ) 
\nonumber \\ [1ex] 
&\;&\;\;\;\;\;\;\;\;\;\;\;\;\;\;\;\;\; \times (X_a^{\;2}+P_a^{\;2}) 
\nonumber \\ [1ex]  
&=&\delta_{ab}\frac{2}{N\Gamma_N}
\int\mbox{d}\Omega_{2N}\int_0^\infty\mbox{d}R\; R^{2N-1}
\nonumber \\ [1ex] 
&\;&\;\;\;\;\;\;\;\;\;\;\;\;\;\;\;\;\; \times\delta\Big (R+\sqrt 2)(R-\sqrt 2)\Big )  
\nonumber \\ [1ex] \label{surfaceeval} 
&=&\delta_{ab}
\;\;, \end{eqnarray} 
making use of isotropy and, in particular,
replacing $X_a^{\;2}+P_a^{\;2}$ by 
$\sum_{a'}(X_{a'}^{\;2}+P_{a'}^{\;2})/N=2/N$ under the integral; in the end, 
we employ $2N$-dimensional spherical coordinates, where $\Omega_{2N}$ denotes 
the spherical angle, and choose the normalization factor appropriately:  
\begin{equation}\label{normalizationfactor} 
\Gamma_N:=\frac{N}{2^{N-1}\Omega_{2N}}=\frac{N!}{(2\pi )^N}
\;\;. \end{equation} 
Thus, we obtain from Eq.\,(\ref{CLreduced1}) the expected simple result: 
\begin{equation}\label{CLreduced2}
\rho_{\mbox{\scriptsize CL}}(x_k,p_k)
=\sum_jw_j(x_k,p_k)
\;\;, \end{equation}
using completeness and orthonormality of the bases. 

\subsection{Quantum-classical Poisson bracket and separability}
The result of the calculation in Eq.\,(\ref{rhotransf}) suggests to introduce a  
{\it generalized Poisson bracket}, when considering 
observables defined on the Cartesian product state space of CL {\it and} QM 
sectors as follows: 
\begin{eqnarray}\label{GenPoissonBracket} 
\{ A,B\}_\times &:=&\{ A,B\}_{\mbox{\scriptsize CL}}+\{ A,B\}_{\mbox{\scriptsize QM}}
\\ [1ex] 
&:=&\sum_k\Big (\frac{\partial A}{\partial_{x_k}}\frac{\partial B}{\partial_{p_k}}
-\frac{\partial A}{\partial_{p_k}}\frac{\partial B}{\partial_{x_k}}\Big )
\nonumber \\ [1ex]  \label{GenPoissonBracketdef}  
&\;&+\sum_i\Big (\frac{\partial A}{\partial_{X_i}}\frac{\partial B}{\partial_{P_i}}
-\frac{\partial A}{\partial_{P_i}}\frac{\partial B}{\partial_{X_i}}\Big ) 
\;\;, \end{eqnarray} 
for any two observables $A,B$. It is bilinear and antisymmetric, leads to a derivation-like 
product formula and obeys the Jacobi identity, since the right-hand side of 
Eq.\,(\ref{GenPoissonBracketdef}) can be written in standard form as a single sum, 
after relabeling the canonical coordinates. 

Let us say {\it an observable ``belongs'' to the CL (QM) sector, if it is  
constant with respect to the canonical coordinates of the QM (CL) sector}. -- 
Then, the generalized Poisson bracket has the additional important properties: 
\begin{itemize} 
\item It reduces to the Poisson brackets introduced 
in Eqs.\,(\ref{PoissonBracket}) and (\ref{QMPoissonBracket}), respectively,   
for pairs of observables that belong {\it either} to the CL {\it or} the QM sector. 
\item It reduces to the appropriate one of the former brackets, 
if one of the observables belongs only to either one of the two sectors. 
\item It reflects the {\it separability} of CL and QM sectors, 
since $\{ A,B\}_\times =0$, if $A$ and $B$ belong to different sectors. 
\end{itemize}  

The physical relevance of separability can be expressed as the following requirement: 
{\it If a canonical tranformation 
is performed on the QM (CL) sector only, then all observables that belong to the 
CL (QM) sector should remain unaffected.}
This is indeed the case, as we shall demonstrate directly by 
examining the behaviour of the reduced CL (QM) probability distribution under 
such transformations. 

Performing in the QM sector, for example, an infinitesimal canonical transformation on the 
integral of Eq.\,(\ref{CLreduced}), we obtain: 
\begin{eqnarray}
&\;&\rho_{\mbox{\scriptsize CL}}(x_k,p_k)\;\rightarrow\; 
\nonumber \\ [1ex] 
&\;&\Gamma_N^{-1}\int_{\delta S_{2N}'(\sqrt 2)}\Pi_j(\mbox{d}X_j'\mbox{d}P_j')\;
\rho (x_k,p_k;X_i',P_i') \;=\;
\nonumber \\ [1ex] 
&\;&\int 
\rho (x_k,p_k;X_i+\frac{\partial G(X_i,P_i)}{\partial P_i}\delta\alpha,
P_i-\frac{\partial G(X_i,P_i)}{\partial X_i}\delta\alpha) 
\nonumber \\ [1ex] 
&\;&=\int
\rho (x_k,p_k;X_i,P_i)  
\nonumber \\ [1ex] 
&\;&\;\;\;+\int 
(\partial_{X_i}\rho\;\partial_{P_i}G-\partial_{P_i}\rho\;\partial_{X_i}G)\delta\alpha 
+\mbox{O}(\delta\alpha^2) 
\nonumber \\ [1ex] \label{CLreducedTrans} 
&\;&=\;\rho_{\mbox{\scriptsize CL}}(x_k,p_k)+\mbox{O}(\delta\alpha^2) 
\;\;, \end{eqnarray} 
where we abbreviated 
$\Gamma_N^{-1}\int_{\delta S_{2N}(\sqrt 2)}\Pi_j(\mbox{d}X_j\mbox{d}P_j)\equiv\int$ 
and used the well known invariance of the phase space volume element and of 
the constraint surface, by Eq.\,(\ref{constraintinvar1}), together with  
Eqs.\,(\ref{Xcan})--(\ref{goperator}); furthermore, the last equality follows 
from the fact that the {\it integral of a Poisson bracket}  
of observables over QM state space vanishes: 
\begin{eqnarray}
&\;&\;\;\;\;\int\;
\{ A,B\}_{\mbox{\scriptsize QM}} =\int\;
\langle\Psi |\frac{1}{i}[\hat A,\hat B]|\Psi\rangle 
\nonumber \\ [1ex] \label{PoissonInt} 
&\;&=\int\;
\mbox{Tr}(|\Psi\rangle\langle\Psi |\frac{1}{i}[\hat A,\hat B]) = 
\mbox{Tr}(\frac{1}{i}[\hat A,\hat B])=0 
\;,\;\;\;  \end{eqnarray}
using Eq.\,(\ref{QMPBComm}), followed by a calculation similar to the one leading 
from Eq.\,(\ref{rhointerpr}) to Eq.\,(\ref{CLreduced2}), via Eqs.\,(\ref{CLreduced}) and 
Eqs.\,(\ref{CLreduced1})--(\ref{normalizationfactor}). -- 
In the present case, incidentally, we have that 
$\{ A,B\}_{\mbox{\scriptsize QM}}\equiv\{\rho ,G\}_{\mbox{\scriptsize QM}}=\{\rho ,G\}_\times$, 
since $G$ belongs to the QM sector. 

Thus, we find invariance of $\rho_{\mbox{\scriptsize CL}}$ under infinitesimal and, hence, finite 
canonical transformations in the QM sector. Consequently, the  
expectation of any CL observable $g_{\mbox{\scriptsize CL}}$, defined by:
\begin{equation}\label{CLexpect}  
\langle g_{\mbox{\scriptsize CL}}\rangle 
:=\int\Pi_l(\mbox{d}x_l\mbox{d}p_l)\; 
g_{\mbox{\scriptsize CL}}(x_k,p_k)\rho_{\mbox{\scriptsize CL}}(x_k,p_k) 
\;\;, \end{equation}  
is invariant. -- 
Similarly, one shows that $\rho_{\mbox{\scriptsize QM}}$ is invariant under 
canonical transformations in the CL sector and, thus, expectations of QM observables 
as well. 

Separability, as demonstrated here, has been a crucial issue in  
discussions of earlier attempts to formulate a consistent quantum-classical 
hybrid dynamics, see, for example, Refs.~\cite{CaroSalcedo99,Hall08,Diosi10} and 
references therein.

\subsection{Time evolution of noninteracting quantum-classical ensembles}
In order to illustrate another aspect of the separability of CL and CM sectors, 
as long as there is {\it no} interaction between them, we consider the 
time evolution of the probability distribution generated by 
the total Hamiltonian function ${\cal H}_\Sigma$: 
\begin{equation}\label{Htotal} 
{\cal H}_\Sigma (x_k,p_k;X_i,P_i):={\cal H}_{\mbox{\scriptsize CL}}(x_k,p_k)+
{\cal H}_{\mbox{\scriptsize QM}}(X_i,P_i) 
\;, \end{equation}
where ${\cal H}_{\mbox{\scriptsize CL}}$ denotes an assumed Hamiltonian function for 
the CL sector, while the Hamiltonian function 
${\cal H}_{\mbox{\scriptsize QM}}$ for the QM sector has been 
detailed above, cf. Eqs.\,(\ref{HamiltonianQM1})--(\ref{HamiltonianQM2}). 

Based on Hamilton's equations for both sectors and equipped with the 
generalized Poisson bracket of Eq.\,(\ref{GenPoissonBracket}), we can invoke 
Liouville's theorem to obtain the evolution equation: 
\begin{equation}\label{rhoevol} 
-\partial_t\rho = \{\rho ,{\cal H}_\Sigma\}_\times  
\;\;. \end{equation} 
Clearly, this equation admits a factorizable solution 
$\rho (x_k,p_k;X_i,P_i;t)\equiv\rho (x_k,p_k;t)\cdot\rho (X_i,P_i;t)$, provided 
the initial condition has this property. {\it No spurious correlations}  
are produced by the evolution, which corresponds to 
$\{{\cal H}_{\mbox{\scriptsize CL}},{\cal H}_{\mbox{\scriptsize QM}}\}_\times =0$, by construction. 

In other words, the CL and QM sectors evolve independently, as if the respective 
other sector was absent, and maintain their classical or quantum nature, as long as they 
do not interact. 

\section{Quantum-classical hybrid dynamics}
Following the preparations in Section~3., which concerned quantum-classical composite 
systems, however, without interaction of the CL and QM sectors, we propose here 
the generalization to include such a hypothetical coupling and will study the  
consistency and consequences of such truly hybrid systems.  

This discussion will concern hybrid ensembles, or specific hybrid states, and their dynamics. 
Given the generalized 
Poisson bracket, introduced in Eq.\,(\ref{GenPoissonBracket}), we have to incorporate a hybrid  interaction term ${\cal I}$ in the total Hamiltonian function ${\cal H}_\Sigma$, 
which will serve as the generator of time evolution, as before. 
Therefore, we replace the definition of Eq.\,(\ref{Htotal}), with  
${\cal H}_\Sigma\equiv{\cal H}_\Sigma (x_k,p_k;X_i,P_i)$, by:  
\begin{equation}\label{HtotalInt} 
{\cal H}_\Sigma:={\cal H}_{\mbox{\scriptsize CL}}(x_k,p_k)
+{\cal H}_{\mbox{\scriptsize QM}}(X_i,P_i) 
+{\cal I}(x_k,p_k;X_i,P_i)   
\;. \end{equation}
For ${\cal H}_\Sigma$ to be an {\it observable}, it is necessary that the hybrid interaction  
qualifies as observable, in particular. Further properties of ${\cal H}_\Sigma$ 
will be detailed in due course. 

Then, the evolution 
equation is of the same form as Eq.\,(\ref{rhoevol}): 
$-\partial_t\rho = \{\rho ,{\cal H}_\Sigma\}_\times\;$, where,  
henceforth, the Hamiltonian function includes the interaction term 
${\cal I}$, unless stated otherwise. -- Here, the {\it positive semi-definiteness} 
of $\rho$ holds for the same reason as for the case of the classical Liouville equation, 
namely that the underlying dynamics is described by a Hamiltonian flow.   

\subsection{Energy conservation} 
Having proposed ${\cal H}_\Sigma$ as the generator of time evolution, it also provides the 
natural candidate for the {\it conserved energy} of the hybrid system. 
Since ${\cal H}_\Sigma$ is assumed not to be explicitly time dependent, 
we find, cf. with the general structure of Eq.\,(\ref{evolution}):  
\begin{equation}\label{evolution1}  
\frac{\mbox{d}{\cal H}_\Sigma}{\mbox{d}t}=\{ {\cal H}_\Sigma,{\cal H}_\Sigma \}_\times 
=0 
\;\;, \end{equation} 
an immediate consequence of the antisymmetry of the generalized Poisson bracket. -- Note that 
in the absence of a CL subsystem, this result reduces to the conservation of the  
expectation of the QM Hamilton operator, as it should. More generally, in the absence 
of QM-CL interactions, the classical and quantum mechanical energies simply add.   

\subsection{Generalized Ehrenfest relations for hybrids}
Here we show that the Poisson structure built into the present theory of hybrid 
systems, in particular in the form of underlying Hamiltonian equations of motion, 
translates into generalizations of Ehrenfest relations for coordinate and momentum 
observables. 
  
We consider hybrid systems described by a generic classical {\it Hamiltonian function}  
and a quantum mechanical {\it Hamiltonian operator}, respectively: 
\begin{eqnarray}\label{Hcl}  
{\cal H}_{\mbox{\scriptsize CL}}&:=&\sum_k\frac{p_k^{\; 2}}{2}+v(x_l)  
\;\;, 
\\ [1ex] \label{Hqm} 
\hat H_{\mbox{\scriptsize QM}}&:=&\frac{\hat P^2}{2}+V(\hat X)
\;\;, \end{eqnarray} 
where $v(x_l)\equiv v(x_1,\dots,x_n)$ and $V$ denote relevant potentials, together with a {\it self-adjoint hybrid 
interaction operator} $\hat I(x_k,p_k;\hat X,\hat P)$; note that symmetrical (Weyl) 
ordering is necessary, concerning the noncommuting operators $\hat X$ and $\hat P$. 
We set all masses equal to one here, for simplicity, but will introduce 
them explicitly in the particular case of coupled oscillators below.  
By Eq.\,(\ref{goperator}), this gives rise 
to the following {\it Hamiltonian function} ${\cal H}_\Sigma$: 
\begin{eqnarray}
{\cal H}_\Sigma &=&\sum_k\frac{p_k^{\; 2}}{2}+v(x_l)
+\langle\Psi |\Big (\frac{\hat P^2}{2}+V(\hat x)\Big )|\Psi\rangle 
\nonumber \\ [1ex] 
&\;&
+\langle\Psi |\hat I(x_k,p_k;\hat X,\hat P)|\Psi\rangle 
\nonumber \\ [1ex] 
&=:& {\cal H}_{\mbox{\scriptsize CL}}(x_k,p_k)
+{\cal H}_{\mbox{\scriptsize QM}}(X_i,P_i) 
\nonumber \\ [1ex] \label{HSigmaGen1} 
&\;&
+{\cal I}(x_k,p_k;X_i,P_i)   
\;\;, \end{eqnarray} 
when evaluated in a pure state $|\Psi\rangle$, invoking the oscillator representation  
of Eq.\,(\ref{oscillexp}). Correspondingly, we define {\it coordinate and momentum 
observables}, in the sense of our earlier construction in Section~2.2., 
pertaining to the QM subsystem: 
\begin{equation}\label{QMobservables} 
X(X_i,P_i):=\langle\Psi |\hat X|\Psi\rangle \;\;,\;\;\;
P(X_i,P_i):=\langle\Psi |\hat P|\Psi\rangle 
\;\;. \end{equation}
With these definitions in place, we proceed to determine the equations of motion 
by following the rules of Hamiltonian dynamics. 

The equations of motion for the CL {\it observables} $x_k,p_k$ are: 
\begin{eqnarray}\label{xdot} 
\dot x_k&=&\{ x_k,{\cal H}_\Sigma\}_\times 
=p_k+\partial_{p_k}{\cal I}(x_k,p_k;X_i,P_i)
\;\;, \\ [1ex] 
\dot p_k&=&\{ p_k,{\cal H}_\Sigma\}_\times  
\nonumber \\ [1ex] \label{pdot} 
&=&-\partial_{x_k}v(x_l)-\partial_{x_k}{\cal I}(x_k,p_k;X_i,P_i) 
\;\;. \end{eqnarray} 
Similarly, we obtain for the QM {\it variables} $X_i,P_i$, which are {\it not observables}: 
\begin{eqnarray}
\dot X_i&=&\{ X_i,{\cal H}_\Sigma\}_\times 
\nonumber \\ [1ex] \label{Xidot}  
&=&\partial_{P_i}{\cal H}_{\mbox{\scriptsize QM}}(X_j,P_j)+\partial_{P_i}{\cal I}(x_k,p_k;X_j,P_j)
\\ [1ex] \label{Xidot1} 
&=&E_iP_i+\partial_{P_i}{\cal I}(x_k,p_k;X_j,P_j) 
\;\;, \\ [1ex]  
\dot P_i&=&\{ P_i,{\cal H}_\Sigma\}_\times 
\nonumber \\ [1ex] \label{Pidot}
&=&-\partial_{X_i}{\cal H}_{\mbox{\scriptsize QM}}(X_j,P_j)-\partial_{X_i}{\cal I}(x_k,p_k;X_j,P_j) 
\\ [1ex] \label{Pidot1} 
&=&-E_iX_i-\partial_{X_i}{\cal I}(x_k,p_k;X_j,P_j)
\;\;, \end{eqnarray} 
where Eqs.\,(\ref{Xidot1}) and (\ref{Pidot1}) follow, if the oscillator expansion 
is performed with respect to the stationary states of $\hat H_{\mbox{\scriptsize QM}}$, 
cf. Eqs.\,(\ref{HamiltonianQM1})--(\ref{HamiltonianQM2}) in Subsection~2.2.2. -- 
Notably, the Eqs.\,(\ref{xdot}), (\ref{pdot}) together with Eqs.\,(\ref{Xidot}), (\ref{Pidot}),  
or together with Eqs.\,(\ref{Xidot1}), (\ref{Pidot1}), form a {\it closed set} of $2(n+N)$ 
equations, where $n$ denotes the number of CL degrees of freedom and $N$ the 
dimension of the QM Hilbert space (assumed denumerable, if not finite). 

However, in distinction, the {\it generalized Ehrenfest relations} for the QM 
{\it observables} $X,P$, 
defined in Eqs.\,(\ref{QMobservables}), are obtained as follows: 
\begin{eqnarray}
\dot X&=&\{ X,{\cal H}_\Sigma\}_\times =\{ X,{\cal H}_\Sigma\}_{\mbox{\scriptsize QM}}
\nonumber \\ [1ex] 
&=&-i\langle\Psi |[\hat X,\hat H_{\mbox{\scriptsize QM}}+\hat I]|\Psi\rangle 
\nonumber \\ [1ex] \label{Xdot} 
&=&P
-i\langle\Psi |[\hat X,\hat I(x_k,p_k;\hat X,\hat P)]|\Psi\rangle 
\;\;, \\ [1ex] 
\dot P&=&\{ P,{\cal H}_\Sigma\}_\times =\{ P,{\cal H}_\Sigma\}_{\mbox{\scriptsize QM}}
\nonumber \\ [1ex]
&=&-i\langle\Psi |[\hat P,\hat H_{\mbox{\scriptsize QM}}+\hat I]|\Psi\rangle 
\nonumber \\ [1ex] \label{Pdot}
&=&-\langle\Psi |V'(\hat X)|\Psi\rangle 
-i\langle\Psi |[\hat P,\hat I(x_k,p_k;\hat X,\hat P)]|\Psi\rangle 
,\;\;\;\;\; \end{eqnarray} 
where we used Eq.\,(\ref{QMPBComm}), in order to replace Poisson brackets   
by commutators and the explicit form of $\hat H_{\mbox{\scriptsize QM}}$, Eq.\,(\ref{Hqm}); 
$V'$ denotes the appropriate first derivative of the potential function $V$. -- The 
Eqs.\,(\ref{Xdot})--(\ref{Pdot}) together with Eqs.\,(\ref{xdot})--(\ref{pdot}) do 
{\it not form a closed set of equations}, since the expectation of a function  
of observables generally does not equal the function of the expectations of the observables, 
as in Ehrenfest's theorem in quantum mechanics.~\cite{footnote1} 

\subsection{Bilinearly coupled oscillators}
We consider here a {\it set of} CL oscillators coupled bilinearly 
to {\it one} QM oscillator, choosing, for example: 
\begin{eqnarray}\label{Hcl1}  
{\cal H}_{\mbox{\scriptsize CL}}&:=&\sum_k
\Big ({\textstyle \frac{1}{2m_k}p_k^{\; 2}}
+{\textstyle \frac{m_k\omega_k^{\; 2}}{2}}x_k^{\; 2}\Big ) 
\;\;, 
\\ [1ex] \label{Hqm1} 
\hat H_{\mbox{\scriptsize QM}}&:=&{\textstyle \frac{1}{2M}}\hat P^2
+{\textstyle \frac{M\Omega^2}{2}}\hat X^2 
\;\;, 
\\ [1ex] \label{bilinear} 
\;\;\;\;\;\;\hat I&:=&\hat X\sum_k\lambda_kx_k
\;\;, \end{eqnarray} 
where we introduced masses $m_k,M$, frequencies $\omega_k,\Omega$, 
and coupling constants $\lambda_k$.  

In this case, the equations of motion for the CL observables together with the 
{\it generalized Ehrenfest relations} of the previous subsection 
reduce to a simple {\it closed} set of equations:  
\begin{eqnarray}\label{xdot1} 
\dot x_k&=&{\textstyle \frac{1}{m_k}}p_k
\;\;, \\ [1ex] \label{pdot1} 
\dot p_k&=&-{\textstyle m_k\omega_k^{\; 2}}x_k 
-\lambda_k X 
\;\;, \\ [1ex] \label{Xdot1} 
\dot X&=&{\textstyle \frac{1}{M}}P
\;\;, \\ [1ex] \label{Pdot1}
\dot P&=&-{\textstyle M\Omega^2}X
-\sum_k\lambda_kx_k 
\;\;, \end{eqnarray} 
with the QM observables $X:=\langle\Psi|\hat X|\Psi\rangle$ and 
$P:=\langle\Psi|\hat P|\Psi\rangle$, cf. Eqs.\,(\ref{QMobservables}). Here, the 
{\it backreaction} of QM on CL subsystem appears, {\it as if} 
the CL subsystem was coupled to another CL oscillator.  

In view of Eqs.\,(\ref{xdot1})--(\ref{Pdot1}), we find that our theory 
passes the ``Peres-Terno benchmark'' test for interacting 
QM-CL hybrid systems \cite{PeresTerno}, which, so far, 
has been achieved only by the configuration ensemble theory of Hall and Reginatto 
\cite{HallReginatto05,Hall08,ReginattoHall08}. 

\section{Discussion}
The proposed theory describing QM-CL hybrid systems certainly raises a number of  
questions, some of which we address in the following. 

\subsection{Classical-environment induced decoherence}
Well known studies of environment induced decoherence describe the effects that an  
environment of QM degrees of freedom has on the coherence properties of a QM subsystem 
coupled to it \cite{Kieferetal,Zurek}. In particular, the Feynman-Vernon or Caldeira-Leggett 
models and, more generally, models of quantum Brownian motion have been studied in this 
respect \cite{Grabertetal}. Here we suggest to consider the situation where  
the QM environment is replaced by a classical one. We shall find that a 
CL environment similarly can produce decoherence in a generic model. 
    
For simplicity, we consider a QM object characterized by a two-dimensional 
Hilbert space, a ``q-bit'', which is coupled bilinearly to a set of CL oscillators. 
The oscillators are described by Eq.\,(\ref{Hcl1}), as before. The Hamiltonian 
function of the q-bit presents the simplest example of the oscillator expansion 
of a QM Hamiltonian: 
\begin{equation}\label{2state}
{\cal H}(X_i,P_i):=\sum_{i=1,2}\frac{E_i}{2}(P_i^{\;2}+X_i^{\;2}) 
\;\;, \end{equation} 
when expanding with respect to the energy eigenstates, $\{ |\phi_1\rangle ,|\phi_2\rangle\}$, 
cf. Eqs.\,(\ref{oscillexp})--(\ref{HamiltonianQM2}). 
The bilinear QM-CL coupling is defined by: 
\begin{equation}\label{bilinear1} 
\hat I:=\hat\Sigma\sum_k\lambda_kx_k
\;\;, \end{equation} 
similarly as before in Eq.\,(\ref{bilinear}); here $\hat\Sigma$ denotes 
an observable of the q-bit. 
  
In this case, the closed set of dynamical equations of motion becomes: 
\begin{eqnarray}\label{xdotoscill} 
\dot x_k&=&p_k/m_k
\;\;, \\ [1ex] \label{pdotoscill} 
\dot p_k&=&-m_k\omega_k^{\; 2}x_k-\lambda_k\langle\Psi |\hat\Sigma |\Psi\rangle  
\;\;, \\ [1ex] \label{Xidot1oscill} 
\dot X_i&=&E_iP_i 
\;\;, \\ \label{Pidot1oscill} 
\dot P_i&=&-E_iX_i-\frac{\partial\langle\Psi |\hat\Sigma|\Psi\rangle}{\partial_{X_i}}
\sum_k\lambda_kx_k 
\;\;, \end{eqnarray} 
analogous to Eqs.\,(\ref{xdot}), (\ref{pdot}), (\ref{Xidot1}), (\ref{Pidot1}), and  
where we have: 
\begin{equation}\label{Xoscill}
\langle\Psi |\hat\Sigma |\Psi\rangle =
\frac{1}{2}\sum_{i,j=1,2}\langle\phi_i|\hat\Sigma |\phi_j\rangle (X_i-iP_i)(X_j+iP_j) 
\;. \end{equation} 

The Eqs.\,(\ref{xdotoscill})--(\ref{pdotoscill}) are solved by employing 
the retarded Green's function for the equation of motion of a driven harmonic oscillator. 
This yields: 
\begin{equation}\label{drivenoscill}  
x_k(t)=x_k^{(0)}(t)
-\lambda_k\int_{-\infty}^t\mbox{d}s\;
\frac{\sin\;\omega_k(t-s)}{m_k\omega_k}\langle\Psi (s)|\hat\Sigma |\Psi (s)\rangle  
, \end{equation} 
with the harmonic term $x_k^{(0)}(t):=a_k\cos (\omega_kt)+b_k\sin (\omega_kt)$ and 
where the real coefficients $a_k$ and $b_k$ are determined by initial conditions. 

Furthermore, the Eqs.\,(\ref{Xidot1oscill})--(\ref{Pidot1oscill}) can be 
combined into second order form: 
\begin{eqnarray}
&&\ddot X_i+E_i^{\;2}X_i
\nonumber \\ [1ex] \label{QMoscill}
&&=-E_i\;\xi (t)
\sum_{j=1,2}\Big (\mbox{Re}(\Sigma_{ij})X_j-\mbox{Im}(\Sigma_{ij})\dot X_j/E_j)\Big )
,\;\;\;\;\;\;\;\; \end{eqnarray} 
introducing the matrix elements 
$\Sigma_{ij}:=\langle\phi_i|\hat\Sigma |\phi_j\rangle =\Sigma_{ji}^\ast$, the real and imaginary  parts of which enter. 
Thus, we obtain a system of $N=2$ {\it coupled oscillator equations}, where the coupling 
terms are {\it nonlinear and non-Markovian} through the function: 
\begin{eqnarray} 
\xi (t)&:=&\sum_k\lambda_kx_k(t)
\nonumber \\ [1ex] 
&=&\sum_k\Big [\lambda_kx_k^{(0)}(t)
\nonumber \\ [1ex] 
&\;&-\lambda_k^{\;2}\sum_{i,j=1,2}\Sigma_{ij} 
\int_{-\infty}^t\mbox{d}s\;\frac{\sin\;\omega_k(t-s)}{m_k\omega_k}
\nonumber \\ [1ex] \label{drivenoscill1} 
&\;&\times
\Big (X_i(s)X_j(s)+\dot X_i(s)\dot X_j(s)/(E_iE_j)\Big )\Big ]
\;,\;\;\;\; \end{eqnarray} 
by Eqs.\,(\ref{Xidot1oscill}), (\ref{Xoscill})-(\ref{drivenoscill}).  

Let us reduce the above model to a crudely simplified version, neglecting presumably much 
of the rich dynamics described by Eqs.\,(\ref{QMoscill})--(\ref{drivenoscill1}). -- 
For sufficiently {\it weak coupling}, we drop the non-Markovian terms, {\it i.e.},  
terms $\propto\lambda_k^{\;2}$. Furthermore, we choose $\Sigma_{11}=\Sigma_{22}\equiv 0$ 
and $\Sigma_{12}=\Sigma_{21}\equiv 1$. This simplifies the equations to 
describe two oscillators which are symmetrically 
coupled to each other through a periodic or, in the case of CL 
oscillators with incommensurate frequencies, quasi-periodic function $\xi$. -- 
Under the additional assumption of {\it slow CL oscillators}, {\it i.e.}, with frequencies 
that are small compared to the ones of the QM oscillators, the resulting equations 
are solved by $(i=1,2)$: 
\begin{equation}\label{QMsol} 
X_i=A_i\cos (\Omega_it)+B_i\sin (\Omega_it)
\;\;, \end{equation} 
and $P_i=\dot X_i/E_i$; the real coefficients $A_i,\;B_i$ are 
determined by initial conditions and, to leading non-vanishing order in $\xi$,   
the characteristic frequencies are given by: 
\begin{equation}\label{QMfrequ} 
\Omega_{1}:=E_1+\frac{\xi^2E_2}{2(E_1^{\; 2}-E_2^{\; 2})}\;,\;\; 
\Omega_{2}:=E_2-\frac{\xi^2E_1}{2(E_1^{\; 2}-E_2^{\; 2})}
\;. \end{equation} 

Choosing, for illustration, initial conditions such that the expansion coefficients 
in $|\Psi\rangle =\sum_{1,2}|\phi_i\rangle (X_i+iP_i)/\sqrt 2$ (cf. Eq.\,(\ref{oscillexp})) 
are real at $t=0$, we obtain the off-diagonal matrix elements of the corresponding 
density matrix $\hat\rho :=|\Psi\rangle\langle\Psi |$ in the form: 
\begin{eqnarray}\label{densmatrix} 
&&\langle\phi_1|\hat\rho |\phi_2\rangle =\langle\phi_2|\hat\rho |\phi_1\rangle^\ast 
=(X_1+iP_1)(X_2-iP_2)/2
\nonumber \\ [1ex]
&&=\;e^{i(\Omega_2-\Omega_1)t}(1-\frac{\xi^2}{4E_1E_2})
\nonumber \\ [1ex] 
&&\;\;\;\;-\frac{\xi^2}{4(E_1^{\; 2}-E_2^{\; 2})}(
\frac{E_2}{E_1}e^{i(\Omega_1+\Omega_2)t}
-\frac{E_1}{E_2}e^{-i(\Omega_1+\Omega_2)t})
\nonumber \\ [1ex] \label{QMdensmatrix} 
&&\;\;\;\;+\mbox{O}(\xi^4)
\;\;. \end{eqnarray} 
We note that there is a term $\propto i\xi^2t$ contributing to the argument of each 
exponential, cf. Eqs.\,(\ref{QMfrequ}). This can lead to 
{\it decoherence by dephasing} in the following way. 

If the nonnegative function $\xi^2(t)$ is sufficiently irregular (depending 
on the frequency distribution of environmental oscillators), we may 
treat it as a random variable and average the result of Eq.\,(\ref{QMdensmatrix}) 
correspondingly. We consider the leading term, while the others can similarly be dealt with. 
Thus, writing $\Omega_2-\Omega_1=\delta E+\xi^2/2\delta E$, with $\delta E:=E_2-E_1$, 
we have to evaluate the dimensionless function $f$: 
\begin{equation}\label{average} 
f(t):=\int_0^\infty\mbox{d}\Omega\; P(\Omega )e^{i\Omega t}
\;\;, \end{equation}
with $\Omega\equiv\xi^2/2\delta E$, and where $P$ represents the appropriately 
normalized distribution of the values of $\Omega$. 
Now, there are {\it continuous distributions}, 
such that $f(t)\rightarrow 0$, for $t\rightarrow\infty$; for example, a constant distribution 
over a finite range of $\Omega$, an exponential distribution, or a Gaussian 
distribution. Under these circumstances, the leading term 
(similarly the others) gives 
a decaying contribution, {\it i.e.} $\propto f(t)$, to the off-diagonal density matrix 
element $\langle\phi_1|\hat\rho |\phi_2\rangle$, after averaging. 

This indicates a decoherence mechanism which is effectively quite similar to 
``fundamental energy decoherence'', which has been reviewed recently 
in Ref.\,\cite{DiosiRev05}.   

\subsection{Quantum-classical backreaction}
Quantum-classical backreaction, in particular the effect of quantum fluctuations 
on the classical subsystem, has always been an important topic for various proposals 
of quantum-classical hybrid dynamics and its applications. This concerns improvements of 
approximation methods and applications, for example, in ``semiclassical gravity'' 
studying the effects of quantum fluctuations of matter on the classical metric 
of spacetime; see Refs.\,\cite{BoucherTraschen,HallReginatto05,ReginattoHall08,HuetalGravity08} 
with numerous references to related work. 

Our formalism consistently {\it incorporates all quantum fluctuations}, 
even if they are not 
explicitly visible, unlike in many approaches where fluctuations are added by hand, in some 
approximation. Presently, the quantum dynamics is treated exactly in terms of 
a complete set of canonical variables, for example, $(X_i,P_i)_{i=1,\dots ,N}$ 
in the closed 
set of dynamical equations (\ref{xdot})--(\ref{Pidot1}). As long as no approximations 
are applied to these equations, their solutions allow to evaluate exactly all  
quantities which reflect the fluctuations associated with a pure quantum state 
$|\Psi\rangle$, such as the typical variance 
$\Delta X^2:=\langle\Psi |\hat X^2|\Psi\rangle -\langle\Psi |\hat X|\Psi\rangle ^2$. This 
follows from the fact that {\it all quantum observables} can be expanded in the oscillator 
representation, recall Eqs.\,(\ref{oscillexp}), (\ref{goperator}), (\ref{Gexp}), 
with the expansion coefficients given by the solutions of the deterministic equations. 
Thus, for example, $\Delta X^2$ becomes a function of the 
canonical variables. 

The QM variables do not fluctuate in a given pure state. By the QM-CL Poisson brackets 
and ensuing equations of motion (``Hamilton's equations'') they are  
coupled to the CL variables (observables) which, therefore,   
do not show fluctuations either. 

However, this admits the possibility that 
the initial conditions of the hybrid dynamical equations, in particular for 
the QM subsystem, are determined by the fluctuating outcome of a certain preparation / 
measurement. In this case, if only statistical / conditional information is available 
about the initial state of the system, the classical observables, generally, will reflect 
corresponding fluctuations. For example, we can evaluate a correlation function 
of CL observables to find:  
$$\langle x_ax_b\rangle:=
\int\Pi_l(\mbox{d}x_l\mbox{d}p_l)\; x_ax_b\rho_{\mbox{\scriptsize CL}}(x_k,p_k)
\neq\langle x_a\rangle\langle x_b\rangle\;,$$   
with the help of the reduced distribution $\rho_{\mbox{\scriptsize CL}}$ 
introduced in Eq.\,(\ref{CLreduced}). This distribution function 
is determined by the solution of the Liouville 
equation for the full density $\rho$ of the interacting hybrid system, cf. Eq.\,(\ref{rhoevol});   
it could be, furthermore, conditioned by a selected outcome of some quantum measurement(s)  
specifying the initial state.  

\subsection{Hybrid observables, separable interactions and QM-CL Poisson brackets} 
It is a common feature of either QM or CL systems that particular forms of interaction among 
subsystems allow to separate degrees of freedom into noninteracting subsets. Generally, 
this is associated with the existence of {\it symmetries} of the compound system, such as 
translation or rotation invariance.  

A recent study of a translation invariant harmonic interaction between a QM and a CL 
particle reveiled that -- according to the hybrid theory proposed by 
Hall and Reginatto -- there arises an irreducible coupling between 
center-of-mass and relative motion \cite{HallReginatto05}. 
This is contrary to what happens if both particles are treated as either classical 
or quantum mechanical and has been traced to the inherent nonlinearity of 
their proposal. The action functional, from which the equations of motion are 
derived, ``knows'' which variables belong to the QM and CL sectors, respectively, 
and mixing them by coordinate transformations produces the coupling. 
Consequently, such a system has been proposed as a prospective testing ground, 
where their theory could be falsified experimentally \cite{HallEtAll11}.   

This issue can also be examined in the light of the present linear hybrid theory. -- 
Specializing the system of bilinearly coupled 
oscillators of Subsection~4.3. as follows: 
\begin{equation}\label{transoscill} 
{\cal H}_{\mbox{\scriptsize CL}}:={\textstyle \frac{1}{2m}p^{\; 2}}, \;\;
\hat H_{\mbox{\scriptsize QM}}:={\textstyle \frac{1}{2M}}\hat P^2, \;\;
\hat I:=\lambda (x\cdot\hat{\bm1}-\hat X)^2 
, \end{equation} 
we reconsider the example of Ref.\,\cite{HallEtAll11}; 
here $\hat{\bm 1}$ denotes the unit operator on the Hilbert space of the QM 
subsystem. 

As before, the Hamiltonian function which generates time evolution of the composite system, 
${\cal H}_\Sigma :={\cal H}_{\mbox{\scriptsize CL}}
+\langle\Psi |(\hat H_{\mbox{\scriptsize QM}}+\hat I)|\Psi\rangle$, 
is conserved by construction, $\mbox{d}{\cal H}_\Sigma /\mbox{d}t=0$, see Subsection~4.1. 
Furthermore, we know from Subsection~4.3. that the generalized Ehrenfest equations 
for bilinearly coupled oscillators in terms of the CL observables, here $x$ and $p$, and 
of the QM observables, $X:=\langle\Psi |\hat X|\Psi\rangle$ and 
$P:=\langle\Psi |\hat P|\Psi\rangle$, form a closed set, cf. Eqs.\,(\ref{xdot1})--(\ref{Pdot1}). 
These equations of motion are nothing but Hamilton's equations for the ``classical''  
Hamiltonian function: 
\begin{equation}\label{classHamiltonian} 
{\cal H}_\Sigma^{\mbox{cl}}(x,p;X,P):=
{\textstyle \frac{1}{2m}p^{\; 2}}+{\textstyle \frac{1}{2M}}P^2+
\lambda (x-X)^2  
\;\;, \end{equation}  
which implies that also ${\cal H}_\Sigma^{\mbox{cl}}$ is conserved, 
$\mbox{d}{\cal H}_\Sigma^{\mbox{cl}} /\mbox{d}t=0$. Then, it follows that the  
energy carried by quantum fluctuations is separately conserved:  
\begin{eqnarray}
&&\frac{\mbox{d}}{\mbox{d}t}\Big (
{\textstyle \frac{1}{2M}}(\langle\hat P^2\rangle -\langle\hat P\rangle^2)+
\lambda (\langle\hat X^2\rangle -\langle\hat X\rangle^2) 
\Big ) 
\nonumber \\ [1ex] \label{fluctenergy} 
&&=\frac{\mbox{d}}{\mbox{d}t}({\cal H}_\Sigma-{\cal H}_\Sigma^{\mbox{cl}}) 
=0
\;\;, \end{eqnarray}
where $\langle \dots\rangle\equiv\langle\Psi |\dots |\Psi\rangle$. 
Note that all mixed QM-CL terms cancel and no energy is transferred 
between the (fluctuations of the) QM and the CL subsystem.  

We observe that the ``classical'' Hamiltonian ${\cal H}_\Sigma^{\mbox{cl}}$ is  
separable by transforming the variables $x,p$ and $X,P$ to center-of-mass and relative 
variables: 
\begin{eqnarray}\label{com} 
&\;&s:=(MX+mx)/\sigma\;\;,\;\;\; p_s:=P+p 
\;\;, 
\\ [1ex] \label{relative} 
&\;&r:=X-x\;\;,\;\;\; p_r:=\mu ({\textstyle \frac{1}{M}}P-{\textstyle \frac{1}{m}}p) 
\;\;, \end{eqnarray} 
with the total and reduced masses defined by $\sigma :=M+m$ and $\mu :=Mm/(M+m)$, 
respectively. Thus, we find: 
${\cal H}_\Sigma^{\mbox{cl}}=(p_s^{\; 2}/2\sigma )+(p_r^{\; 2}/2\mu )+\lambda r^2$, 
not surprisingly.  

At this point, it seems worth while to assess the character of these transformations with 
respect to the fundamental QM-CL Poisson brackets, defined in 
Eqs.\,(\ref{GenPoissonBracket})--(\ref{GenPoissonBracketdef}) (Subsection~3.2.), 
on which our theory is based. We have $\{ x,p\}_\times =1$ and 
$\{ X,P\}_\times =\langle [\hat X,\hat P]/i\rangle =1$, using Eq.\,(\ref{QMPBComm}); 
similarly, we find that the brackets of all other pairs of these variables 
vanish. Thus, we may consider $x,p$ and $X,P$ as two pairs of canonical phase space 
coordinates.  
Furthermore, one may check that also the two pairs of 
center-of-mass and relative variables, $s,p_s$ and $r,p_r$, respectively, form 
pairs of canonical coordinates under the QM-CL Poisson brackets. 
Therefore, the transformations (\ref{com})--(\ref{relative}) are consistent 
{\it canonical transformations}.     

An immediate consequence is that the ``classical Hamiltonians'' describing center-of-mass 
and relative motion are separately conserved as well. 

It must be emphasized that the 
separation of the ``classical'' degrees of freedom, $x,p$ and $X,P$, or of the corresponding 
center-of-mass and relative variables, from the full set of canonical variables $x,p$ and 
$X_i,P_i$, is an {\it accident of the harmonic interaction}. Independently of the hybrid 
coupling between a classical and a quantum mechanical particle, cf. Eqs.\,(\ref{transoscill}), 
any other but constant, linear (in one dimension), or harmonic translation invariant coupling 
would not allow such 
separation, even if both particles were treated quantum mechanically. 

In general, the 
``mean field'' variables 
$X\equiv\langle\hat X\rangle$ and $P\equiv\langle\hat P\rangle$ will always 
couple to ``correlation functions'', such as $\langle\hat X^2\rangle$, 
$\langle\hat X\hat P+\hat P\hat X\rangle$, $\langle\hat P^2\rangle$, or more complicated ones, 
depending on the kind of interaction. This phenomenon of quantum mechanics is not 
particular to hybrid dynamics. 

While the separation of degrees of freedom in the case of translation or 
rotation invariant potentials in quantum mechanics can always be completed  
at the operator level, the hybrid dynamics presented here 
necessitates the consideration of ``classical'' canonical variables on which 
to perform any canonical transformations consistently with the underlying Poisson bracket 
structure. This seems to limit separability to certain potentials, as we have just seen. 

We conclude that a composite system of a QM and a CL particle with 
harmonic translation invariant interaction, or some analogue of this, 
does not allow to experimentally falsify our formulation of hybrid dynamics. 
We find no coupling between relative and 
center-of-mass motion, contrary to the proposal of Refs.\,\cite{HallReginatto05,HallEtAll11}. 
However, anharmonic 
interactions need to be studied in this context and may  
lead to experimentally accessible signatures of the linear QM-CL hybrid dynamics. 

\subsection{The classical$\;\times\;$almost-classical algebra of hybrid observables}
In Subsections~2.1. and 2.2., we introduced the notions of classical and quantum 
observables, respectively, relevant for the considerations of this paper. 

Furthermore, in Subsection~3.2., we introduced the fundamental QM-CL hybrid Poisson bracket, 
$\{ A,B\}_\times :=\{ A,B\}_{\mbox{\scriptsize CL}}+\{ A,B\}_{\mbox{\scriptsize QM}}$. 
Following Eqs.\,(\ref{GenPoissonBracket})--(\ref{GenPoissonBracketdef}), we pointed out 
three important properties of this bracket, last not least related to separability.  
However, we tacitly assumed that a fourth case would not need further mention, 
which can arise for two genuine hybrid observables:  
\begin{itemize}
\item Let $A\equiv A(x_k,p_k;X_i,P_i)$, $B\equiv B(x_k,p_k;X_i,P_i)$ 
be {\it both  
hybrid observables}, i.e., both are quadratic forms in the $X_i$'s and $P_i$'s and   
both are not completely independent of the $x_k$'s and $p_k$'s. If, furthermore, 
one observable, say $A$, depends on any pair of canonical variables, say $x\equiv x_{k'}$ and 
$p\equiv p_{k'}$, and $B$ also depends on $x$ or $p$, then 
the ``classical part'' of the bracket, $\{ A,B\}_{\mbox{\scriptsize CL}}$, 
generates terms which do {\it not} qualify as observable with respect to the QM sector. 
\end{itemize}
Such terms are of the general form: 
\begin{eqnarray} 
&&\sum_{i,i',j,j'}M_{i,j,i',j'}(x_k,p_k)
(X_i-iP_i)(X_j+iP_j)
\nonumber \\ [1ex]
&&\;\;\;\;\;\;\;\;\times (X_{i'}-iP_{i'})(X_{j'}+iP_{j'}) 
\nonumber \\ [1ex] 
&&=4\sum_{i,i',j,j'}\langle\Psi |\Phi_i\rangle\langle\Psi |\Phi_{i'}\rangle 
M_{i,j,i',j'}(x_k,p_k)
\nonumber \\ [1ex] \label{nonlinear} 
&&\;\;\;\;\;\;\;\;\times \langle\Phi_j|\Psi\rangle\langle\Phi_{j'}|\Psi\rangle
\;\;, \end{eqnarray} 
where we used the oscillator expansion, Eq.\,({\ref{oscillexp}), and: 
$$M_{i,j,i',j'}(x_k,p_k):=\frac{1}{4}\sum_k
\Big (\frac{\partial A_{ij}}{\partial_{x_k}}\frac{\partial B_{i'j'}}{\partial_{p_k}}
-\frac{\partial A_{ij}}{\partial_{p_k}}\frac{\partial B_{i'j'}}{\partial_{x_k}}\Big )
, $$ 
using the related expansion for observables $A$ and $B$, cf. Eq.\,(\ref{Gexp}). 

Generally, iterations of such brackets will implicitly contribute    
to the solution, $\rho\equiv\rho (x_k,p_k;X_i,P_i)$, of the evolution equation, 
$-\partial_t\rho = \{\rho ,{\cal H}_\Sigma\}_\times\;$, in the presence of 
a true hybrid coupling, cf. Section~4. Thus, multiple factors involving 
the state vector $|\Psi\rangle$ and its adjoint, or multiple pairs 
like $(X_i-iP_i)(X_j+iP_j)$, will enter. In this way,   
evolution of hybrid observables, of the density $\rho$ 
in particular, can induce a structural change: while continuing to be CL 
observables, they do not remain QM observables (quadratic forms in $X_i$'s and $P_i$'s). 
They fall outside of the product algebra generated by the 
observables to which we confined ourselves, so far. 

We note that the assumption of a 
product algebra covering the observables of a hybrid system was essential   
for the no-go theorem put forth in Ref.\,\cite{Salcedo96}, 
which ruled out a class of hybridization models. However, 
this assumption can be criticized as being too  
restrictive from the QM point of view \cite{Hall08}. -- 
Here we assume: 
\begin{itemize} 
\item The algebra of hybrid 
observables is closed under the QM-CL Poisson bracket $\{\;,\;\}_\times$ 
operation -- a physical hypothesis. 
\end{itemize} 
 
Refering to the phase space coordinates $(X_i,P_i)$, 
we define an {\it almost-classical observable} as a real-valued regular function 
of pairs of factors like $(X_i-iP_i)(X_j+iP_j)$, such as in the 
left-hand side of Eq.\,(\ref{nonlinear}), subject to the constraint:   
${\cal C}(X_i,P_i)=\frac{1}{2}\sum_i(X_i^{\;2}+P_i^{\;2})\stackrel{!}{=}1$. 
This normalization constraint, cf. Eq.\,(\ref{oscillnormalization}) in Subsection~2.2.2., 
is preserved under the evolution, since $\{{\cal C},{\cal H}_\Sigma\}_\times =0$, 
in the presence of QM-CL hybrid interaction as defined in Section~4.  

According to this definition, QM observables 
(quadratic forms in phase space coordinates, cf. Section~2.2.) 
form a subset of almost-classical observables which, in turn, form a subset of 
classical observables 
(real-valued regular functions of phase space coordinates, cf. Section~2.1.).  

Furthermore, we may now say that members of the complete algebra of hybrid observables, 
generally, are {\it classical} with respect to 
coordinates $(x_k,p_k)$ and {\it almost-classical} with respect to coordinates $(X_i,P_i)$.  

This leads us to speculate about a physical consequence of the enlarged 
classical$\;\times\;$almost-classical algebra for interacting QM-CL hybrids, 
as illustrated by the following {\it Gedankenexperiment}. 

Consider a quantum together with a classical object subject to a transient hybrid interaction. 
As long as the hybrid interaction is ineffective, both objects 
evolve independently according to Schr\"odinger's and Hamilton's equations, respectively. 
However, once they form an interacting hybrid, the corresponding 
phase space density changes from a factorized form, in absence of any initial 
correlation, to become an 
almost-classical/classical hybrid observable. Even if the hybrid interaction eventually 
disappears, the density possibly maintains a mixed almost-classical/classical character. 
This agrees with the general structure of the evolution equation, yet  
needs to be understood in detailed examples. 

This outcome contradicts naive expectation that quantum and classical objects 
evolve separately in quantum and classical ways, {\it after} any hybrid interaction 
has ceased. -- Two possibilities come to mind. Either persistence of the 
almost-classical/classical character is a {\it physical 
effect} accompanying QM-CL hybrids, if they exist. Or our description  
needs to be augmented with a {\it reduction mechanism} by which evolving observables return to  
standard QM or CL form (cf. Section~2.), following a hybrid interaction.  
Both possibilities seem quite interesting in their own right. We reserve this topic 
for future study.~\cite{footnote2} 

\subsection{Hybrid dynamics and Wigner function approach}
Suppose a physicist unfamiliar with quantum mechanics were presented with the 
general equations of motion, Eqs.\,(\ref{xdot})--(\ref{Pidot1}) (plus normalization constraint, 
Eq.\,(\ref{oscillnormalization})). -- {\it We know} that these equations present independent CL 
and QM sectors, in the absence of a hybrid interaction. -- However, he/she would naturally 
interpret them to describe the dynamics of a 
composite CL object, with part of its phase space compactified, due to the constraint. 
Thus, he/she finds {\it perfectly local dynamics}.  
In fact, {\it our} knowledge of nonlocal features can be traced to the definition of   
the canonical coordinates and momenta $X_i,P_i$, introduced by the oscillator representation, 
Eq.\,(\ref{oscillexp}), since: $X_i/\sqrt 2=\mbox{Re}\int\mbox{d}q\;\Phi^\ast _i(q)\Psi(q)$ 
and $P_i/\sqrt 2=\mbox{Im}\int\mbox{d}q\;\Phi^\ast _i(q)\Psi(q)$. Therefore, 
spatially nonlocal (and probabilistic) features enter by reference to the 
QM wave function.~\cite{footnote3} 

In view of this, it might be surprising that our proposed hybrid dynamics passes the 
set of consistency requirements, cf. Section~1., in particular the requirement of conservation 
and positivity of probability, as we have seen. 

This must be contrasted with the problems 
that arise if one maps the QM sector ``locally'' on a would-be classical phase space by using the 
Wigner function approach and the corresponding version of the von\,Neumann equation. 
The latter differs from the classical Liouville equation by a series of corrections in powers of 
$\hbar$  which, in turn, incorporate nonlocal features.  
It is well known that  they  spoil the interpretation of the Wigner function as 
a genuine probability distribution on phase space, since it generally does not remain positive 
semidefinite, see Ref.\,\cite{Mueckenheim} for a comprehensive review on probability 
issues.   

Conversely, the dynamics of classical phase space distributions, typically described by 
the Liouville equation, can be presented in quantum mechanical and, in particular, in 
path integral form \cite{EGV11,EGV10}. Again, the resulting would-be quantum mechanical 
density matrix corresponding to a classical probability distribution is, generally, not 
positive semidefinite.  

In both cases, the problem is caused by the intermediate Fourier transformation, which 
apparently is not suited to represent the nonlocality properties appropriately, 
when formally relating phase space to Hilbert space and {\it vice versa}. Therefore, the Wigner 
function approach, which one could be tempted to employ, in order to systematically reduce 
part of a composite QM system to a CL subsystem, thus defining a QM-CL hybrid, 
unfortunately violates the positivity of probability requirement.   

Similar problems have been encountered in Ref.\,\cite{DiosiGisinStrunz}, where the 
QM$\rightarrow$CL reduction is attempted via coherent  ``minimum uncertainty'' states. 

 In distinction, 
the oscillator representation allows to circumvent this difficulty, at the expense of  
introducing the phase space structure  in an abstract way.~\cite{footnote4}

\section{Concluding remarks} 
We have proposed a theory of {\it quantum-classical hybrid dynamics} in this paper. 
In particular, our considerations are based on the representation of quantum mechanics 
in the framework of classical analytical mechanics by Heslot, who showed that 
notions of states in phase space, observables, and Poisson brackets can be 
naturally extended to quantum mechanics \cite{Heslot85}. 

Our formulation provides 
a generalization for the case, where quantum mechanical and classical
degrees of freedom are directly coupled to each other. An important guideline has been 
to satisfy the complete set of {\it consistency conditions} mentioned in Section~1. 
and fulfilled, so far, only by the 
configuration space ensemble theory of Hall and 
Reginatto \cite{HallReginatto05,Hall08,ReginattoHall08}, while all earlier attempts 
failed in one or the other point. 
However, our linear theory deviates from their nonlinear theory in that no `spurious' 
coupling between center-of-mass and relative motion is 
found for a two-body system with a harmonic translation invariant potential.   

This latter issue, quantum-classical backreaction, classical-environment induced 
decoherence, and completion of the algebra of hybrid observables have been discussed in 
Section~5., while further interesting topics are 
left for future work. These include: the hypothetical role of hybrid dynamics in measurement 
processes, seen as the interaction between a classical apparatus and a quantum object 
according to the Copenhagen interpretation, and the effect of classical and quantum  
degrees of freedom on entangled and classically correlated states, respectively, through hybrid 
interactions. 

On the technical side, since hybrid dynamics in the present formulation leads to 
a Liouville equation, as we discussed, the superspace path integral we have recently 
devised can be readily adapted to it \cite{EGV11,EGV10}. 
This may be particularly interesting for applications in 
which hybrid dynamics is considered as an approximation scheme for complex quantum systems.  

In a more speculative vein, 
one could wonder about the essential difference between quantum and classical state 
spaces seen here, respectively, in the presence and absence of curvature (cf. Section~2.2.2.). 
Does a properly understood classical limit of quantum mechanics, which possibly helps 
with the measurement problem~\cite{footnote5}, 
with a commented extensive list of references given in the last one.}, require a dynamical 
treatment of this curvature? Conversely, is the hypothetical emergence of quantum 
mechanics from deterministic dynamics related to a dynamical structure of phase space?  

\section{Acknowledgements}
It is a pleasure to thank L.~Di\'osi, M.J.~Everitt, F.~Finster, V.I.~Man'ko, 
T.M.~Nieuwenhuizen, and T.~Padmanabhan for discussions, 
A.~Khrennikov and S.~Burra for invitations to present related work in  
lectures at ``Foundations of Probability and Physics - 6''  
(V$\ddot{\mbox{a}}$xj$\ddot{\mbox{o}}$) 
and at ``Frontiers in Fundamental Physics - 12'' (Udine), 
respectively, J.~Clemente-Gallardo, H.D. Liu, L.L.~Salcedo, 
and M.J.W. Hall and M. Reginatto for very helpful correspondence. 



\begin{thebibliography}{99}
\bibitem{footnote3} Further discussion of locality, entanglement, and  ``Free Will'' 
in hybrid dynamics can be found in \cite{Elze12}.
\bibitem{footnote5} Useful discussions of the measurement 
problem from representative and differing points of view can be found in 
\cite{Adler03,GianCarlo09,Nieuwenhuizen}.
\end{thebibliography}
\end{document}